\documentclass[12pt,a4paper,fleqn]{article}

\usepackage[english]{babel}
\usepackage{amssymb}
\usepackage{amsmath}
\usepackage{txfonts}
\usepackage{mathdots}
\usepackage{geometry}
\usepackage[classicReIm]{kpfonts}
\usepackage[marginal]{footmisc}

\usepackage{graphicx}
\usepackage{titlesec}
\usepackage{setspace}
\usepackage{paralist}
\usepackage{enumitem}
\usepackage{xcolor}
\usepackage{framed}
\usepackage{lipsum}
\usepackage{fancypar}
\usepackage{adjustbox}
\usepackage{CJKutf8}
\usepackage{multirow}
\usepackage{arydshln}
\usepackage{colortbl}

\definecolor{shadecolor}{rgb}{0.92,0.92,0.92}
\titleformat*{\section}{\large\bfseries}
\newcommand{\retract}{\hangafter=1 \setlength{\hangindent}{1.2em}}

\begin{document}

{\centering\Large{\textbf{Chinese E-Romance: Analyzing and Visualizing 7.92 Million Alibaba Valentine's Day Purchases}}}

\noindent \textbf{}

\noindent \textbf{Yongzhen Wang\textsuperscript{1,*}, Xiaozhong Liu\textsuperscript{2,*}, Yingnan Ju\textsuperscript{3}, Katy B\"{o}rner\textsuperscript{3}, Jun Lin\textsuperscript{4}, Changlong Sun\textsuperscript{4}, Luo Si\textsuperscript{4}}

\noindent \textbf{}

\noindent \textbf{1 WISE Lab, Institute of Science of Science and S\&T Management, Faculty of Humanities and Social Sciences, Dalian University of Technology, Dalian 116024, China\\ 2 Computer Science Department, Worcester Polytechnic Institute, Worcester, MA 01609, USA\\ 3 School of Informatics, Computing and Engineering, Indiana University, Bloomington, IN 47408, USA\\ 4 Alibaba DAMO Academy, Hangzhou 310030, China\\ * yongzhenwang@dlut.edu.cn (YW)\quad xliu14@wpi.edu (XL)}

\noindent \textbf{}

\noindent \textbf{Abstract}: The days that precede Valentine's Day are characterized by extensive gift shopping activities all across the globe. In China, where much shopping takes place online, there has been an explosive growth in e-commerce sales during Valentine's Day over the recent years. This exploratory study investigates the extent to which each product category and each shopper group can exhibit romantic love within China's e-market throughout the 2 weeks leading up to 2019 Valentine's Day. Massive data from Alibaba, the biggest e-commerce retailer worldwide, are utilized to formulate an innovative romance index (RI) to quantitatively measure e-romantic values for products and shoppers. On this basis, millions of shoppers, along with their millions of products purchased around Valentine's Day, are analyzed as a case study to demonstrate their love consumption and romantic gift-giving. The results of the analysis are then illustrated to help understand Chinese e-romance based on the perspectives of different product categories and shopper groups. This empirical information visualization also contributes to improving the segmentation, targeting, and positioning of China's e-market for Valentine's Day.\\


\section{Introduction}

\noindent Valentine's Day on February 14$^{\rm th}$ is a perfect time for lovers to express their affection with intimacies and gifts. In China, Valentine’s Day has become an increasingly popular date for romantic business, and Chinese lovebirds on this day of the year are becoming even more generous than their Western counterparts. According to the research by MasterCard and National Retail Federation, Chinese consumers spent an average of \$274 on 2018 Valentine’s Day versus \$144 for American ones (Wang \& Tung, 2019). In addition, the previous study also indicates that compared with Westerners, individuals embedded in East Asian cultures are more likely to use gift-giving as a mode of expressing romantic love (Liang \& Murshed, 2015). China, the world's top retail market with an estimated sales volume of \$5.6 trillion in 2019 (Davidson, 2019), provides a natural arena to investigate love consumption and romantic gift-giving on Valentine's Day.\\

\noindent Along with the e-commerce era's progress, gift-givers hoping to woo their love interests have been turning to online shopping venues. Between 2013 and 2014, e-commerce sales during Valentine’s Day experienced a sharp increase of 8\% based on the IBM Digital Analytics Benchmark (Heller, 2014). More recently, the National Retail Federation released data that indicate e-commerce sales around Valentine's Day had boomed in 2018 when 29\% of shoppers made their festival purchases via online stores --- an approximate 2\% annual growth (Wolinsky, 2019). In China, where 35.3\% of the retail sales occur online (Clark, 2019), all major e-commerce platforms show a roaring demand for Valentine's Day gifts. For example, Tmall.com launched about 500,000 new products from 20,000 brands ahead of 2019 Valentine’s Day, including customized items, limited-edition lipsticks, and gift boxes from various suppliers such as Givenchy, Jo Malone, Longines, and Adidas. As another example, throughout the time leading up to 2019 Valentine’s Day, the gross merchandise volume of fresh flowers on Taobao.com soared by 69\% year-on-year, especially with roses increasing to 220\%. These observations demonstrate not only the extensive popularity of Valentine's Day in China's e-market but also the commercial value of mining China's e-commerce trends in love consumption and romantic gift-giving. Furthermore, while considerable consumer research is focused on expressing emotions in general (Gaur, Herjanto, \& Makkar, 2014), studies on conveying romantic love are surprisingly sparse in marketing literature.\\

\noindent In this study, our goal is to explore Chinese e-romance from a practical point of view, which attempts to understand the love consumption and romantic gift-giving of China's online shoppers by analyzing the massive Valentine's Day purchase-data on Alibaba for 2019. Alibaba, the biggest e-commerce company worldwide, is leading China's e-market with 53.3\% of the online retail sales and 758 million active users (Blystone, 2021). The latest market survey by KPMG and Mei.com suggests that Alibaba is the most popular e-commerce platform in China, with 44\% of the millennials choosing it as their favorite online shopping channel (Mehra, 2017). For these reasons, we select Alibaba as our testbed to probe into the extent to which each product category and each shopper group can exhibit romantic love within China's e-market over the Valentine's Day period. First, we tracked 1.07 million shoppers and their 7.92 million products purchased around Valentine's Day. Next, we formulated an innovative romance index (RI) to measure products' and shoppers' e-romantic values quantitatively. Finally, we visualized the RI analysis results to help illustrate Chinese e-romance from the perspectives of different product categories and shopper groups. In addition, this empirical information visualization also contributes to improving the segmentation, targeting, and positioning of China's market for Valentine's Day.\\

\section{Literature Review}

\noindent In this section, we briefly review the studies on love consumption and romantic gift-giving in literature.\\

\noindent Along with booming online commercial events, gift shopping has become a critical factor in improving e-commerce sales (Ye, Gai, Youssef, \& Jiang, 2019). Valentine's Day, featuring many deals and promotions, is an important occasion of gift shopping for consumers who are enthusiastic about expressing romantic love. One of the typical researches on love consumption was that of Close and Zinkhan (2006), who analyzed Valentine's rituals, themes, and meanings to understand consumer behavior for this holiday. With further research, Close and Zinkhan (2009) aimed to uncover the attitude and behavior of anti-consumption and alternative consumption on Valentine's Day to advance resistance theories. Recently, Zayas, Pandey, and Tabak (2017) indicated that as Valentine’s Day neared, both roses and chocolates would be evaluated more positively in the United States. Moreover, they investigated how attachment style influenced love consumption across various contexts, from movies, books, and greeting cards to romantic gaming and online dating (Mende, Scott, Garvey, \& Bolton, 2019). Despite these successful practices, few studies have been devoted explicitly to an exploration of love consumption within China's retail market.\\

\noindent The importance of gift-giving lies in its ability to help forge and reinforce emotional bonds between givers and recipients (Belk \& Coon, 1993). This notion is particularly true for intimate relationships, where gifts can serve as signals to convey romantic love, caring, and trust (Cheal, 1987). One of the early researches on romantic gift-giving stemmed from a study by Netemeyer, Andrews, and Durvasula (1993), which compared three behavioral intention models for Valentine's Day gift-giving in terms of their capabilities to predict and explain business intelligence. Later, Otnes, Ruth, and Milbourne (1994) examined the questions of what males believed about the purpose of Valentine's Day, what they liked the most and least about this day, and why they did or did not participate in gift-giving activities. Similarly, Rugimbana Donahay, Neal, and Polonsky (2003) made attempts to figure out the motivation of gift-giving by young males on Valentine's Day and showed that individual motivation might have deeper manifestations in the perceived social power relationship. More recently, Lai and Huang (2013) explored consumer decision to purchase fresh flowers as Valentine’s Day gifts based on relationship stage, affection, and satisfaction with the relationship. In general, the majority of existing studies focus on Western cultures, leaving many questions about China's romantic gift-giving unanswered.\\

\noindent In this study, the proposed approach leverages massive data pertaining to Alibaba Valentine's Day purchases to investigate the love consumption and romantic gift-giving in China. To our knowledge, we make the first attempt to address e-commerce romance by utilizing extensive purchase records, and the practice of romantically indexing products and shoppers in e-commerce environments, followed in this paper, is also less explored. In addition, this study is launched on a large-scale real-world dataset with up to 1.07 million Alibaba users to make the research findings more convincing.

\section{Methodology}

\noindent \textbf{2.1{\qquad}Data Collection}

\noindent The dataset used in this study includes 1.07 million Alibaba users' 7.92 million purchase records throughout the 2 weeks between 02-01-2019 and 02-14-2019. Meanwhile, 10.07 million purchase records from the same Alibaba users over 14 random ordinary shopping days of 2019 are collected for comparisons. All selected users are active, i.e., each of them made at least three purchases in 2019. For each user, we can retrieve and determine six of the individual's demographic features, including gender, age, birthplace, occupation, income, and residence, as summarized in Table 1. Here, the gender, age, and birthplace are parsed from user registration information; the occupation is self-reported by users; the income and residence are two Alibaba-dependent ratings inferred by algorithms based on payment history and shipping address, respectively. Providing more detailed information, we go on to state that there are a total of 707,349 female and 362,196 male users, and their age distribution is 371,838 in 15$\sim$24, 394,305 in 25$\sim$34, 216,146 in 35$\sim$44, 72,637 in 45$\sim$54, and 14,619 in 55$\sim$64. Note that those users who are under 15 years or above 64 years of age are excluded from this study. In addition, after de-duplicating all available purchase records, we obtain 6.38 million unique products in total.\\

\noindent Since Alibaba takes concerns of data privacy seriously, we would like to emphasize that none of the purchase records in this study's database permits specific identification with a particular individual, and that the database retains no information about the identity, IP address, or specific physical location of any user.\\

\newpage

\noindent Table 1

\noindent The Demographic Features of Our Dataset

\noindent\begin{tabular}{cp{5cm}p{7cm}}
        \hline
        \textbf{Feature Name} & \textbf{Collection Method} & \textbf{Description}\\ \hline
        gender & parsed from registration information & female and male \\
        age & parsed from registration information & 5 age groups, including 15$\sim$24, 25$\sim$34, 35$\sim$44, 45$\sim$54, and 55$\sim$64 \\
        birthplace & parsed from registration information & 31 provinces of China mainland, including Anhui, Beijing, Chongqing, Fujian, Gansu, Guangdong, Guangxi, Guizhou, Hainan, Hebei, Heilongjiang, Henan, Hubei, Hunan, Inner Mongolia, Jiangsu, Jiangxi, Jilin, Liaoning, Ningxia, Qinghai, Shaanxi, Shandong, Shanghai, Shanxi, Sichuan, Tianjin, Tibet, Xinjiang, Yunnan, and Zhejiang \\
        occupation & self-reported by users & 14 employment types, including agriculture/forestry/animal husbandry/fishery, business service, culture/sports/entertainment, education, financial service, freelance/other service, government, information technology, junior faculty, manufacturing, medical/health service, scientific/technical service, student, and transit/transport service \\
        income & inferred by algorithms & 5 levels from low to high: 1$\sim$5 \\
        residence & inferred by algorithms & 6 levels from urban to rural: 1$\sim$6 \\ \hline
      \end{tabular}\label{tab:1} \\\\

\noindent \textbf{2.2{\qquad}Data Analysis}

\noindent First of all, we seek to formalize the definition of ``{\it e-romantic products}'' around Valentine's Day. In China, the terminology for gifts is \begin{CJK*}{UTF8}{gbsn}礼物\end{CJK*}, in which the first token \begin{CJK*}{UTF8}{gbsn}礼\end{CJK*} refers to the expectation of following a rule of social interactions (Ye, Gai, Youssef, \& Jiang, 2019). According to this implication, we retrieve all products containing keywords about romantic relationships (e.g., lover, couple, boyfriend/girlfriend, and associated synonyms) in their titles, descriptions, or customer reviews as the candidates for research. This practice considerably eliminates those purchase records irrelevant to love consumption and romantic gift-giving. On the other hand, it alleviates the impact of some other online commercial events on Valentine's Day purchases to a significant degree. For instance, the 2 weeks ahead of 2019 Valentine's Day cover the Chinese Spring Festival on February 5$^{\rm th}$, and this festival may also inspire gift-giving activities among families, friends, and relatives.\\

\noindent Next, we can calculate the RI score for a specific product, as shown in Eq. (1), where $I\in\mathbb{N}^+$ denotes the number of total products; ${\rm PV}(\mathcal{P}_i)\in[0,1]$ and ${\rm PO}(\mathcal{P}_i)\in[0,1]$ represent two probabilities that a shopper buys the product $\mathcal{P}_i$ within the 2 weeks ahead of 2019 Valentine's Day and within the 14 random ordinary shopping days, respectively. Both probabilities are estimated by dividing the number of purchase records within the corresponding period by the number of total shoppers. In this manner, a higher ${\rm RI}(\mathcal{P}_i)\in\mathbb{R}^+$ indicates that $\mathcal{P}_i$ gains more popularity over the 2019 Valentine's Day period than over the ordinary shopping period and consequently has a closer correlation to Chinese e-romance. In practice, only those products having RI scores above 1.20 are defined as e-romantic products; this indicates that the sales volume per capita is at least 20\% higher within the 2 weeks ahead of 2019 Valentine's Day than within the 14 random ordinary shopping days. Note that the threshold 1.20 was determined empirically. As for the non-e-romantic products, we simply specify their RI scores to 1.00 for convenience. On this basis, we will be able to compute the RI score for a particular shopper, as shown in Eq. (2), where $J\in\mathbb{N}^+$ denotes the number of total shoppers; $\vec{\mathcal E}_j=[\mathcal{E}_{j1},\ \mathcal{E}_{j2},\ \cdots,\ \mathcal{E}_{jI}]$ represents the purchase record of the shopper $\mathcal{S}_j$ around Valentine's Day; ${\rm RI}(\vec{\mathcal P})=[{\rm RI}(\mathcal{P}_1),\ {\rm RI}(\mathcal{P}_2),\ \cdots,\ {\rm RI}(\mathcal{P}_I)]$ indicates the RI scores of all products; and $\otimes$ refers to the Hadamard multiplication. More concretely, $\mathcal{E}_{ji}=1$ if $\mathcal{S}_j$ has bought $\mathcal{P}_i$, otherwise $\mathcal{E}_{ji}=0$. Note that we choose the statistical function Max$(\cdot)$ instead of Mean$(\cdot)$ because most shoppers pick up one major gift rather than multiple minor ones for significant others.
\begin{equation}
{\rm RI}(\mathcal{P}_i)=\dfrac{{\rm PV}(\mathcal{P}_i)}{{\rm PO}(\mathcal{P}_i)}\quad\forall i\in\{1,2,\cdots,I\} 
\end{equation}
\begin{equation}
{\rm RI}(\mathcal{S}_j)={\rm Max}\big(\vec{\mathcal E}_j\otimes{\rm RI}(\vec{\mathcal P})\big)\quad\forall j\in\{1,2,\cdots,J\}
\end{equation}

\noindent Both kinds of RI scores are then normalized between 1 and 100, with 1 denoting the least e-romantic and 100 representing the most e-romantic. However, a low RI score does not necessarily mean that either a product or a shopper is not romantic on all occasions, and it merely indicates that a product or a shopper has a limited impact on 2019 Valentine's Day at Alibaba. Fig. 1 illustrates the flow diagram of our data analysis, in which the numbers of available products, candidate products, and e-romantic products are presented, individually. We observe that a total of 15,212 e-romantic products are finalized, and they account for less than 1\% of the available ones. We also find that only 9.04\% of the users in the dataset have bought an e-romantic product over the 2019 Valentine's Day period via Alibaba services and define them as ``{\it e-romantic shoppers}'' accordingly. Although the definitions of e-romantic products and e-romantic shoppers can miss some true love consumption and romantic gift-giving, they greatly support the empirical analysis for Chinese e-romance with the help of million Valentine's Day purchases.\\

\noindent \includegraphics[height=115mm]{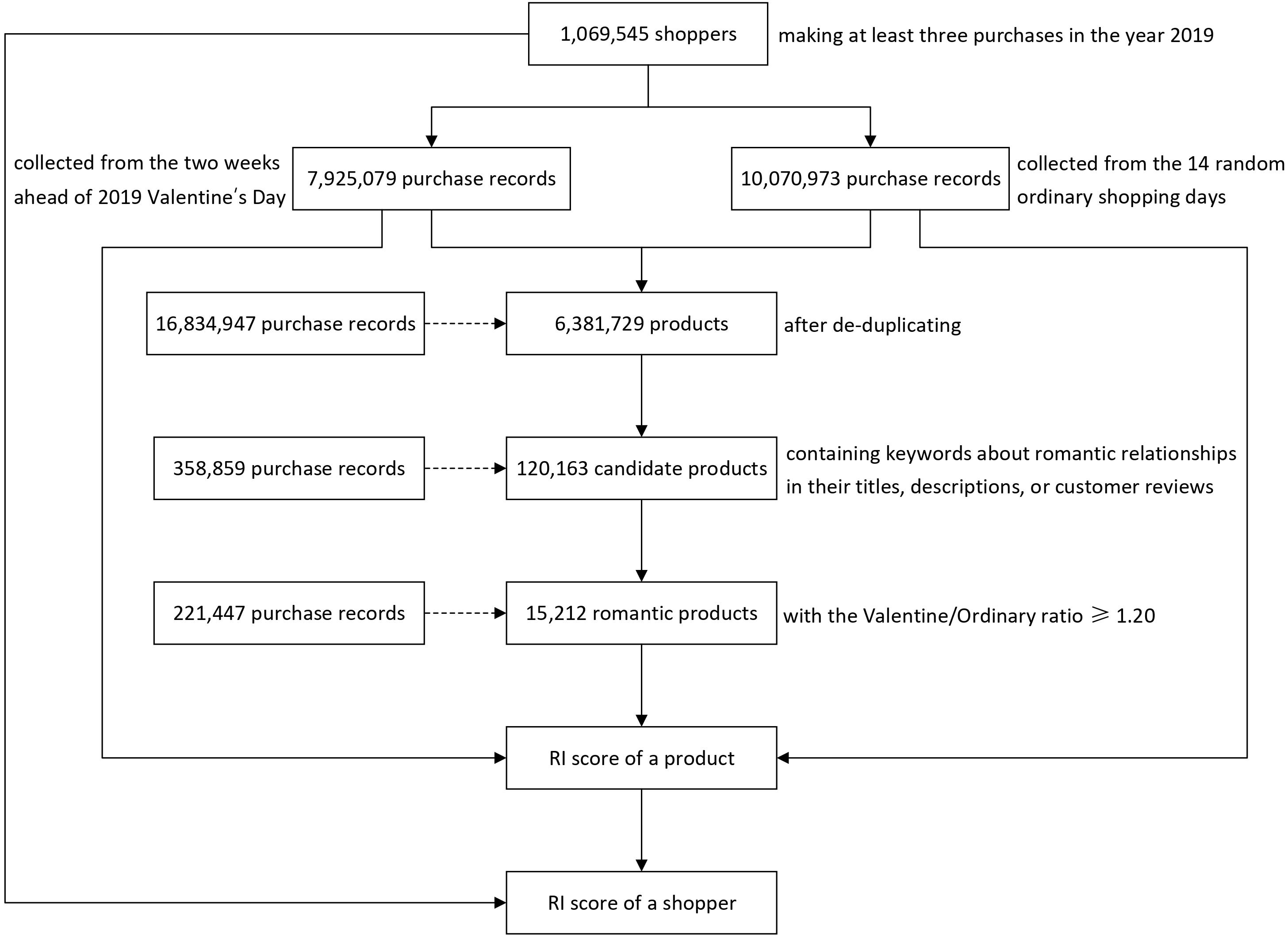}

\noindent \textbf{}

\noindent \textit{Figure 1.} The flow diagram of our data analysis. RI, romance index

\section{Results and Visualization}

\noindent Fig. 2 visualizes the analysis results of e-romantic products in terms of sales volume per capita and associated sales trends throughout the 2 weeks leading up to 2019 Valentine's Day at Alibaba. Note that large and small-sized circles indicate high and low sales volumes per capita, respectively; the rise and fall of sales trends per capita are depicted as well. All e-romantic products are organized into five product categories, and these are gift, sex, apparel, d$\acute{e}$cor, and beauty. For instance, apparel-oriented items consist of pants, shoes, clothes, hats, and pajamas. \\

\noindent Considering all 2019 Valentine's Day gifts sold via Alibaba, gift-oriented items are the top buying option, whereas beauty-oriented gifts are the least popular. Except for d$\acute{e}$cor and beauty, the other product categories start increasing 9$\sim$10 days ahead of 2019 Valentine's Day and reach their peaks 2$\sim$3 days before that. As might be expected, female shoppers make more purchases than their male counterparts during 2019 Valentine's Day at Alibaba, which is consistent with the common understanding that women are more deeply involved in gift shopping activities than men. Meanwhile, there is a gender difference that female shoppers buy more apparel-oriented items, while male shoppers buy more sex-oriented ones. In addition, a majority of 2019 Valentine's Day purchases at Alibaba occur in the shopper group aged between 15 and 44. One reasonable explanation for this outcome is that young generations may have a greater viscosity to e-commerce sites than older ones due to their broader exposure to internet applications.\\

\noindent Interestingly, compared with jewelry and flower/chocolate (i.e., iconic Valentine's Day gifts), both daily (e.g., mugs and electric toothbrushes) and creative/DIY (e.g., personalized cellphone cases and customized puzzles) oriented items report a higher sales volume per capita over the 2019 Valentine's Day period at Alibaba. From the viewpoint of showing romantic love, selecting and buying daily oriented items (especially for personal use) as love consumption often conveys the longing that givers wish to link and live with recipients. Mugs, as one typical example, are a symbol of lifelong engagement in China because the terminology for them is \begin{CJK*}{UTF8}{gbsn}杯子\end{CJK*}, which shares a similar pronunciation to \begin{CJK*}{UTF8}{gbsn}一辈子\end{CJK*} (i.e., a lifetime of love). As for creative/DIY-oriented items, the prior research indicates that consumers' need for uniqueness is increasingly affecting Chinese consumption mode (Ye, Gai, Youssef, \& Jiang, 2019). Under this circumstance, e-commerce platforms play a natural channel for new businesses to sell original and exclusive products not readily available at offline retailers.\\

\noindent \includegraphics[height=88mm]{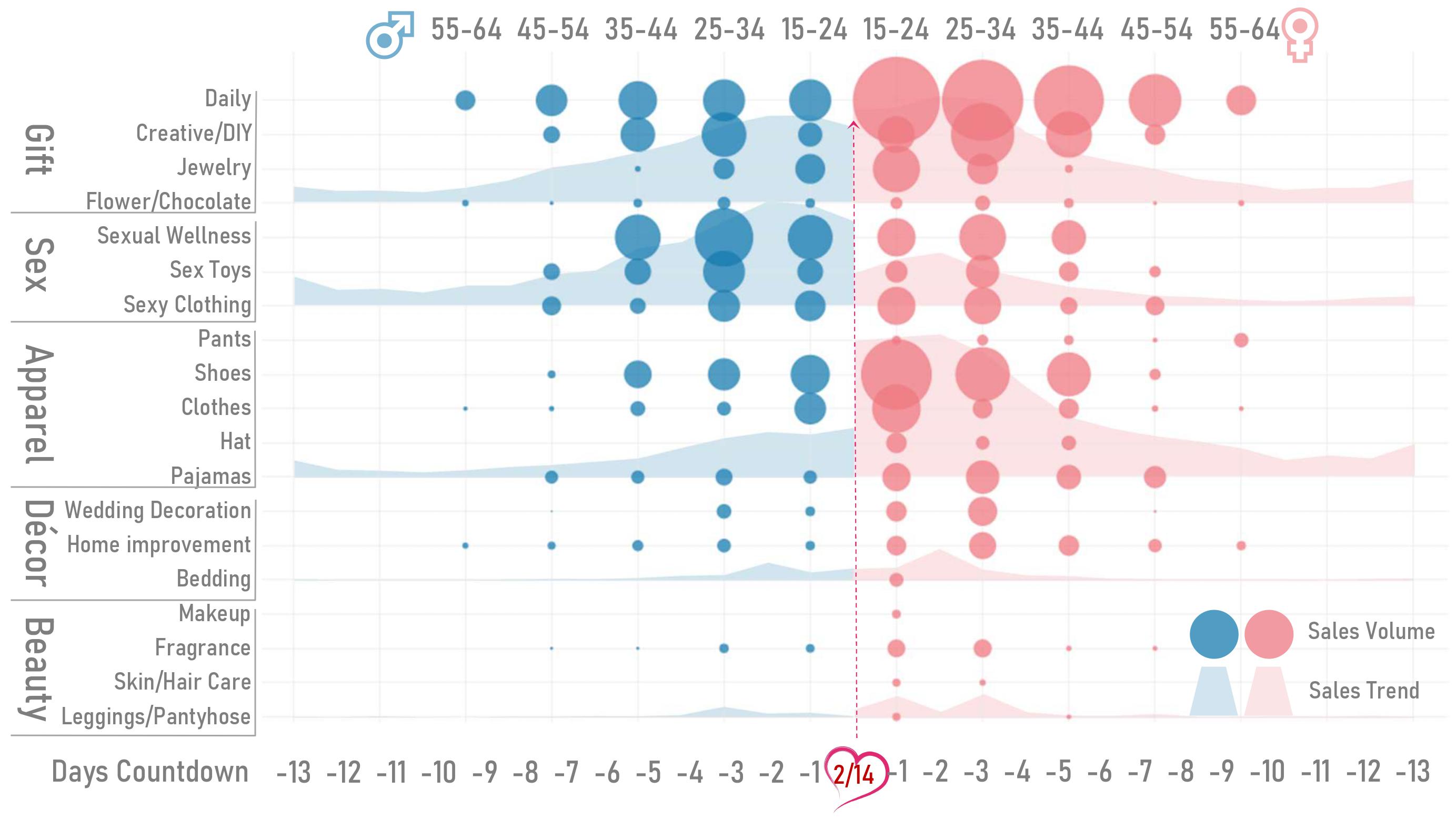}

\noindent \textbf{}

\noindent \textit{Figure 2.} The analysis results of e-romantic products in terms of sales volume per capita and sales trend per capita.

\noindent \textbf{}

\noindent \includegraphics[height=105mm]{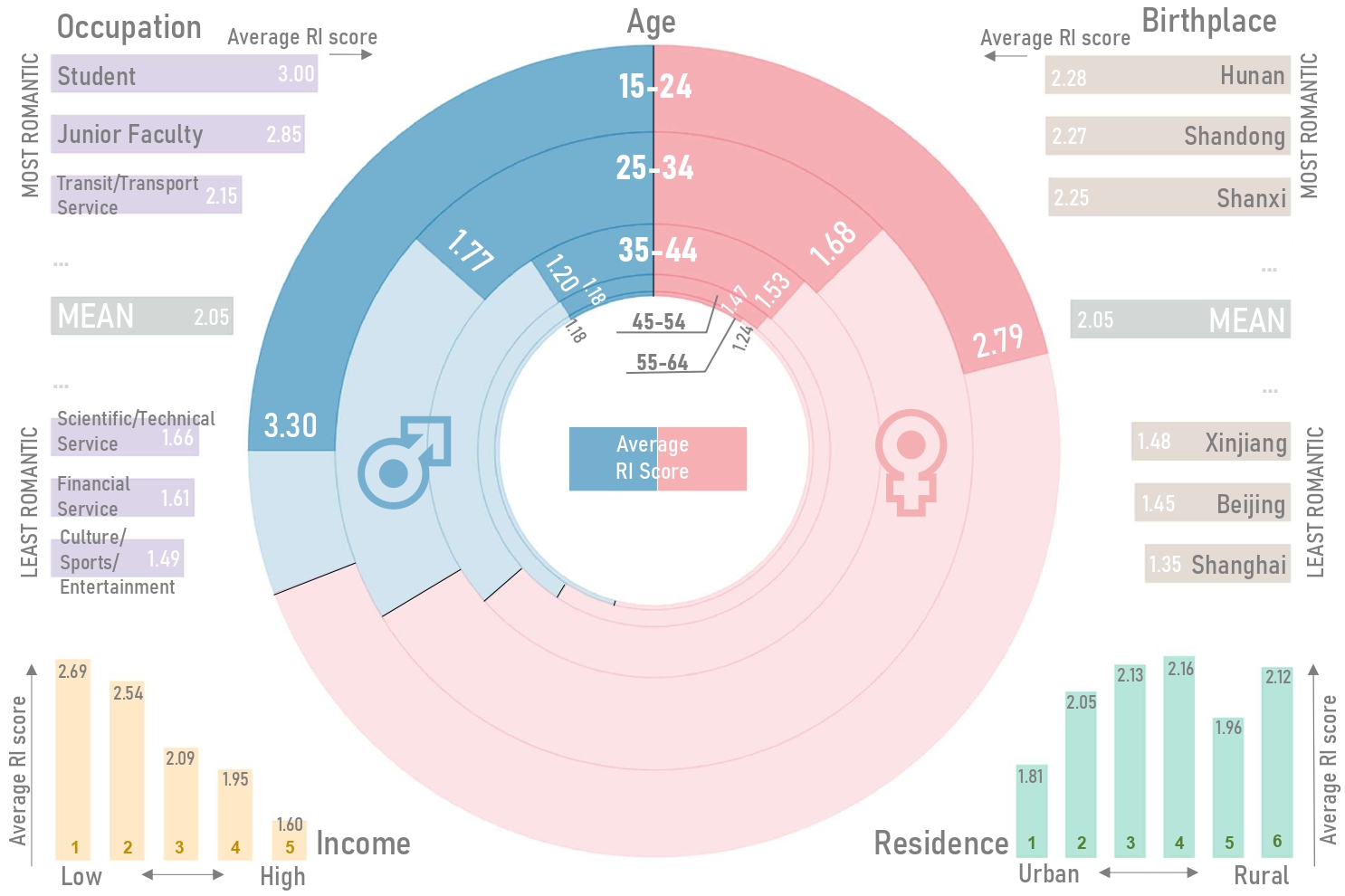}

\noindent \textbf{}

\noindent \textit{Figure 3.} The analysis results of e-romantic shoppers in terms of gender,
age, birthplace, occupation, income, and residence. RI, romance index.

\noindent \textbf{}

\noindent Fig. 3 visualizes the analysis results of e-romantic shoppers in terms of gender, age, birthplace, occupation, income, and residence around 2019 Valentine's Day at Alibaba. Note the following in the middle part --- the wide and narrow diameter of circles indicates the large and small number of shoppers; and the high and low ratio of blue to red is proportional to the gender ratio of males to females. Specifically, our main findings are summarized as follows:\\

\noindent Overall, the average RI score of shoppers decreases with aging, which aligns with the observation in Fig. 2 that the majority of 2019 Valentine's Day purchases at Alibaba are made by younger generations. Furthermore, on average, male shoppers aged 15 to 34 have a higher RI score than their female counterparts, while female shoppers aged 35 to 64 have a higher RI score than the same age male ones. One plausible explanation for this outcome can be marriage status. In China, the average age at first marriage has been postponed to 26 years old (Peng, 2018), and this figure is notably higher for urban areas (e.g., 30 and 28 years old for male and female residents in Shanghai). The recent study suggests that for unmarried Chinese lovers, women not only take greater control over selections of partners but also wield greater decision-making power within love relationships than men, owing to China's imbalanced gender ratio of 48.47\% females to 51.53\% males (Blair \& Madigan, 2016; Ritchie \& Roser, 2019). However, such an initial dominant-inferior setup often changes after marriage as women want love, but men want wives (Singh, 2013).\\

\noindent Table 2

\noindent The RI Scores of Shoppers in Different Occupations

\noindent\begin{tabular}{cp{6cm}ccc}
    \hline
    \multirow{2}*{\bf No} & \multicolumn{2}{c}{\bf Occupation} & \multicolumn{2}{c}{\bf RI Score of Shoppers} \\ \cline{2-5}
    & \bf Name & \bf Annual Wage & \bf Average & \bf Standard Error\\ \hline
    1 & student & N/A & 3.00 & 0.03 \\
    2 & junior faculty & \textyen83,412 & 2.85 & 0.09 \\
    3 & transit/transport service & \textyen80,225 & 2.15 & 0.02 \\
    4 & medical/health service & \textyen89,648 & 2.00 & 0.04 \\
    5 & education & \textyen83,412 & 1.96 & 0.03 \\
    \rowcolor{gray!20}
    6 & freelance/other service & \textyen50,552 & 1.95 & 0.02 \\
    7 & business service & \textyen81,393 & 1.89 & 0.02 \\
    \cdashline{1-5}[0.8pt/2pt]
    8 & information technology & \textyen133,150 & 1.85 & 0.12 \\
    9 & manufacturing & \textyen64,452 & 1.83 & 0.04 \\
    10 & government & \textyen80,372 & 1.75 & 0.04 \\
    \rowcolor{gray!20}
    11 & agriculture/forestry/animal husbandry/fishery & \textyen36,504 & 1.71 & 0.21 \\
    12 & scientific/technical service & \textyen107,815 & 1.66 & 0.15 \\
    13 & financial service & \textyen122,851 & 1.61 & 0.07 \\
    14 & culture/sports/entertainment & \textyen87,803 & 1.49 & 0.13 \\
    \hline
  \end{tabular}\\
  1. The occupations with uncertain and volatile income are marked in grey color. \\
  2. The data on annual wage is referred from the China Statistical Yearbook 2018.\\
  3. RI, romance index.\\\\

\noindent Obviously, different occupations exhibit different average RI scores over the 2019 Valentine's Day period at Alibaba. To elucidate, student is the most e-romantic occupation, followed by junior faculty (e.g., postdoctoral researchers, lecturers, and assistant professors) and transit/transport service, whereas scientific/technical service, financial service, and culture/sports/entertainment are the three least e-romantic ones. More information about the average RI scores of shoppers in different occupations is listed in Table 2. In particular, student and junior faculty contain comparatively younger shoppers than the rest of occupations, which again sustains the observation made for the age distribution of e-romantic shoppers in Fig. 2. After excluding freelance/other service and agriculture/forestry/animal husbandry/fishery, namely the two occupations with uncertain and volatile income, we notice that in the remaining table, the top and bottom halves' average annual wages are \$12,385 (\textyen83,618) and \$14,723 (\textyen99,407), respectively. This outcome, to a certain degree, keeps pace with the correlation between RI and income --- the average RI score of shoppers monotonously decreases as their income levels increase. Possibly, high-income individuals possess more budgets to enjoy brick-and-mortar products and services regarding love consumption and romantic gift-giving.\\

\newpage

\noindent Table 3

\noindent The RI Scores of Shoppers in Different Birthplaces

\noindent\begin{tabular}{ccccc}
    \hline
    \multirow{2}*{\bf No} & \multicolumn{2}{c}{\bf Birthplace} & \multicolumn{2}{c}{\bf RI Score of Shoppers} \\ \cline{2-5}
    & \bf Name & \bf GDP per Capita & \bf Average & \bf Standard Error\\ \hline
    1 & Hunan & \$8,001 & 2.28 & 0.05 \\
    2 & Shandong & \$11,525 & 2.27 & 0.04 \\
    3 & Shanxi & \$6,850 & 2.25 & 0.06 \\
    4 & Hebei & \$7,219 & 2.24 & 0.05 \\
    5 & Jiangxi & \$7,168 & 2.23 & 0.06 \\
    6 & Henan & \$7,579 & 2.22 & 0.04 \\
    7 & Anhui & \$7,210 & 2.17 & 0.04 \\
    8 & Inner Mongolia & \$10,322 & 2.15 & 0.07 \\
    9 & Jilin & \$8,404 & 2.11 & 0.07 \\
    10 & Guangdong & \$13,058 & 2.08 & 0.04 \\
    11 & Liaoning & \$8,766 & 2.06 & 0.05 \\
    12 & Heilongjiang & \$6,539 & 2.05 & 0.06 \\
    13 & Guizhou & \$6,233 & 2.04 & 0.06 \\
    14 & Sichuan & \$7,387 & 2.03 & 0.04 \\
    15 & Shaanxi & \$9,593 & 2.02 & 0.06 \\
    \cdashline{1-5}[0.8pt/2pt]
    16 & Fujian & \$13,781 & 2.00 & 0.05 \\
    17 & Tianjin & \$18,241 & 1.99 & 0.11 \\
    18 & Chongqing & \$9,964 & 1.98 & 0.06 \\
    \rowcolor{gray!20}
    19 & Guangxi & \$6,270 & 1.94 & 0.05 \\
    20 & Yunnan & \$5,612 & 1.90 & 0.06 \\
    21 & Jiangsu & \$17,404 & 1.89 & 0.04 \\
    22 & Zhejiang & \$14,907 & 1.86 & 0.04 \\
    \rowcolor{gray!20}
    23 & Ningxia & \$8,175 & 1.85 & 0.14 \\
    24 & Hainan & \$7,851 & 1.83 & 0.11 \\
    25 & Hubei & \$10,067 & 1.80 & 0.04 \\
    \rowcolor{gray!20}
    26 & Gansu & \$4,735 & 1.71 & 0.06 \\
    27 & Qinghai & \$7,207 & 1.65 & 0.16 \\
    \rowcolor{gray!20}
    28 & Tibet & \$6,564 & 1.53 & 0.34 \\
    \rowcolor{gray!20}
    29 & Xinjiang & \$7,476 & 1.47 & 0.08 \\
    30 & Beijing & \$21,188 & 1.44 & 0.08 \\
    31 & Shanghai & \$20,398 & 1.35 & 0.07 \\
    \hline
  \end{tabular}\\
    1. The five autonomous regions are marked in gray color. \\
    2. The data on GDP per capita is referred from the Statistical Communiqué of the People's Republic of China on the 2018 National Economic and Social Development.\\
    3. RI, romance index.\\\\
    
\newpage

\noindent Surprisingly, among the 31 provinces and municipalities in China mainland, Hunan, Shandong, and Shanxi are the three most e-romantic birthplaces, whereas Xinjiang, Beijing, and Shanghai are the three least e-romantic ones, during the time leading up to 2019 Valentine's Day at Alibaba. More information about the average RI scores of shoppers in different birthplaces is listed in Table 3. Note that five autonomous regions, including Xinjiang, Tibet, Gansu, Ningxia, and Guangxi --- the highest level of minority autonomous entity in China with a comparably higher population of a particular minority ethnic group, all contain shoppers with low RI scores on average. After excluding these regions, we notice that in the remaining table, the top and bottom halves' average GDPs per capita are \$8,390 (\textyen55,520) and \$13,329 (\textyen88,204), respectively. This outcome is roughly consistent with the correlation between RI and residence --- shoppers in urban zones have a lower average RI score than ones in relatively rural zones.\\

\section{Conclusion, Limitations and Implications}

\noindent In this paper, we investigate Chinese e-romance by leveraging massive data pertaining to Alibaba Valentine's Day purchases. To this end, 6,381,729 products and 1,069,545 shoppers are romantically indexed to enable ranking their e-romantic values around Valentine's Day. To our knowledge, we make the first attempt to address e-commerce romance by utilizing extensive purchase records. The analysis results of both e-romantic products and e-romantic shoppers are visualized to help illustrate Chinese e-romance from the perspective of different product categories and shopper groups. In general, the proposed approach not only matches some past findings from existing literature on consumer research but also uncovers many novel phenomena about love consumption and romantic gift-giving in e-commerce environments.\\

\noindent The research findings should be interpreted with caution as this study is limited to Alibaba, one of several major e-commerce platforms in China. Second, Valentine's Day may not be representative of other Chinese occasions when love consumption and romantic gift-giving are made, such as Qixi Festival on July 7$^{\rm th}$ of the Lunar calendar and Single's Day on November 11$^{\rm st}$. Third, this study exclusively covers products containing romantic keywords in their titles, descriptions, or customer reviews; products containing romantic keywords in other text sources are not included. Last but not least, the definitions of e-romantic products and e-romantic shoppers presume that Valentine's Day purchases are made online --- which may not always be the case.\\

\noindent Practically, the research findings will benefit practitioners in China's e-market in the aspects of love consumption and romantic gift-giving by providing them with a unique understanding of market segmentation, targeting, and positioning over Valentine's Day period. Theoretically, these findings will benefit scholars, researchers, and other academics interested in Chinese e-romance by making them familiar with some salient aspects that characterize China's lovebirds' online shopping behavior.\\

\noindent \textbf{Acknowledgments} \\

\noindent Yongzhen Wang was supported by the Fundamental Research Funds for the Central Universities of China, No. DUT21RC(3)068. We thank Alibaba Group for providing this study with the e-commerce data on Valentine's Day purchases, and reviewers and editors for their thoughtful comments and suggestions. \\

\noindent \textbf{References}\\

\begin{small}

\noindent \retract Belk, R. W., \& Coon, G. S. (1993). Gift giving as agapic love: An alternative to the exchange paradigm based on dating experiences.~\textit{The Journal of Consumer Research,}~\textit{20(3)}, 393-417.\\https://doi.org/10.1086/209357

\noindent \retract Blair, S. L., \& Madigan, T. J. (2016). Dating attitudes and expectations among young Chinese adults: An examination of gender differences.~\textit{The Journal of Chinese Sociology,}~\textit{3(1)}, 12.\\https://doi.org/10.1186/s40711-016-0034-1

\noindent \retract Blystone, D. (2021, January 18). \textit{Understanding the Alibaba business model}.\\Retrieved from https://www.investopedia.com/articles/investing/062315/understanding-alibabas-business-model.asp

\noindent \retract Cheal, D. (1987). Showing them you love them: Gift giving and the dialectic of intimacy.~\textit{The Sociological Review,}~\textit{35(1)}, 150-169.\\https://doi.org/10.1111/j.1467-954X.1987.tb00007.x

\noindent \retract Clark, D. (2019, January 23). \textit{2019: China to surpass US in total retail sales}.\\Retrieved from https://www.emarketer.com/newsroom/index.php/2019-china-to-surpass-us-in-total-retail-sales

\noindent \retract Close, A., \& Zinkhan, G. (2006). A holiday loved and loathed: A consumer perspective of Valentine's Day.~\textit{ACR North American Advances,}~\textit{33}, 356-365.\\Retrieved from https://www.acrwebsite.org/volumes/12416/volumes/v33/NA-33

\noindent \retract Close, A. G., \& Zinkhan, G. M. (2009). Market-resistance and Valentine's Day events.~\textit{Journal of Business Research,}~\textit{62(2)}, 200-207.\\https://doi.org/10.1016/j.jbusres.2008.01.027

\noindent \retract Davidson, P. (2019, January 23). \textit{China to top U.S. as world's no. 1 retail market in 2019: Report}.\\Retrieved from https://www.usatoday.com/story/money/2019/01/23/china-top-us-largest-retail-market-2019-report-says/2651447002

\noindent \retract Gaur, S. S., Herjanto, H., \& Makkar, M. (2014). Review of emotions research in marketing, 2002--2013.~\textit{Journal of Retailing and Consumer Services,}~\textit{21(6)}, 917-923.\\https://doi.org/10.1016/j.jretconser.2014.08.009

\noindent \retract Heller, L. (2014, February 18). \textit{Valentine's Day mobile shopping grows 40\% in 2014}.\\Retrieved from https://www.fierceretail.com/operations/valentine-s-day-mobile-shopping-grows-40-2014

\noindent \retract Lai, Y., \& Huang, L. (2013). The effect of relationship characteristics on buying fresh flowers as romantic valentine’s day gifts.~\textit{HortTechnology,}~\textit{23(1)}, 28-37.\\https://doi.org/10.21273/HORTTECH.23.1.28

\noindent \retract Liang, B., \& Murshed, F. (2015). Culture, expressions of romantic love, and gift-giving.~\textit{Journal of International Business Research,}~\textit{14(1)}, 68-84.\\Retrieved from https://search.proquest.com/docview/1693335517?pq-origsite=gscholar~\&fromopenview=true

\noindent \retract Mehra, N. (2017, December 12). \textit{China’s retail sector transforms, millennials are key drivers, finds KPMG and Mei.com survey}.\\Retrieved from https://home.kpmg/cn/en/home/news-media/press-releases/2017/12/china-retail-sector-transforms.html

\noindent \retract Mende, M., Scott, M. L., Garvey, A. M., \& Bolton, L. E. (2019). The marketing of love: How attachment styles affect romantic consumption journeys.~\textit{Journal of the Academy of Marketing Science,}~\textit{47(2)}, 255-273.\\https://doi.org/10.1007/s11747-018-0610-9

\noindent \retract Netemeyer, R. G., Andrews, J. C., \& Durvasula, S. (1993). A comparison of three behavioral intention models: The case of valentine’s day gift-giving.~\textit{ACR North American Advances,}~\textit{20(1)}, 135-141.\\Retrieved from https://www.acrwebsite.org/volumes/7426

\noindent \retract Otnes, C., Ruth, J. A., \& Milbourne, C. C. (1994). The pleasure and pain of being close: Men’s mixed feelings about participation in valentine’s day gift exchange.~\textit{ACR North American Advances,}~\textit{21(1)}, 159-164.\\Retrieved from https://www.acrwebsite.org/volumes/7578/volumes/v21/NA-21/full

\noindent \retract Peng, Y. (2018, March 13). \textit{Across China: Chinese cities see rising marriage age}.\\Retrieved from http://www.xinhuanet.com/english/2018-03/13/c\_137036066.htm

\noindent \retract Ritchie, H. \& Roser, M. (2019). \textit{Gender ratio}.\\Retrieved from https://ourworldindata.org/gender-ratio

\noindent \retract Rugimbana, R., Donahay, B., Neal, C., \& Polonsky, M. J. (2003). The role of social power relations in gift giving on valentine’s day.~\textit{Journal of Consumer Behaviour: An International Research Review,}~\textit{3(1)}, 63-73.\\https://doi.org/10.1002/cb.122

\noindent \retract Singh, S. (2013). Women want love, men want wives: The discourse of romantic love in young adults' future marriage goals.~\textit{Agenda (Durban, South Africa)}~\textit{27(2)}, 22-29.\\https://doi.org/10.1080/10130950.2013.808798

\noindent \retract Wang, J., \& Tung, M. (2019, February 15). \textit{What China is loving this Valentine’s Day}.\\Retrieved from https://www.alizila.com/what-china-is-loving-this-valentines-day

\noindent \retract Wolinsky, J. (2019, February 19). \textit{The top buying trends of Valentine’s Day}.\\Retrieved from https://www.valuewalk.com/2019/02/top-buying-trends-valentines-day

\noindent \retract Ye, L., Gai, L., Youssef, E., \& Jiang, T. (2019). Love consumption at the digital age: Online consumer reviews and romantic gift giving.~\textit{Journal of Global Marketing,}~\textit{32(5)}, 335-355.\\https://doi.org/10.1080/08911762.2018.1564161

\noindent \retract Zayas, V., Pandey, G., \& Tabak, J. (2017). Red roses and gift chocolates are judged more positively in the US near Valentine’s Day: Evidence of naturally occurring cultural priming.~\textit{Frontiers in Psychology,}~\textit{8}, 355.\\https://doi.org/10.3389/fpsyg.2017.00355

\end{small}

\end{document}